\documentclass[fleqn,twoside]{article}
\usepackage{espcrc2}
\usepackage{amsmath}

\usepackage{graphicx}
\title{Quarks stars in SU(2) Nambu-Jona-Lasinio model with vector coupling}

\author{J. G. Coelho\address[MCSD]{Instituto Tecnol\'{o}gico de Aeron\'{a}utica, \\
        Pra\c{c}a Marechal Eduardo Gomes, 50, DCTA, 12228-900, S\~{a}o Jos\'{e} dos Campos, SP, Brasil}%
        \thanks{Acknowledgements: this work was partially supported by FAPESP, CNPq and CAPES/FCT agreement 183/07.},
        C. H. Lenzi\addressmark,
        M. Malheiro\addressmark,
        R. M. Marinho Jr.\addressmark,
        C. Provid\^{e}ncia\address[J]{Centro de F\'{i}sica Computacional, Departamento de F\'{i}sica, Universidade de Coimbra,
        \\
        P-3004-516, Coimbra, Portugal}
        and
        M. Fiolhais\addressmark
        }

\begin{document}

\begin{abstract}
In this work we study the Nambu-Jona-Lasinio model in the SU (2) version with repulsive vector coupling and apply it to
quark stellar matter. We discuss the influence of the vector interaction on the equation of state (EoS) and study quark
stars that are composed of pure quark matter with two flavors. We show that, increasing the vector coupling, we
obtain more massive stars with larger radii for the same central energy density.
\vspace{1pc}
\end{abstract}

\maketitle

The SU(2) version of the Nambu-Jona-Lasinio model (NJL) \cite{klevansky} has been applied to study quark matter and to investigate
quark and hybrid stars \cite{baldo1,buballa1,blaschke,debora}. In this work, we will include a repulsive interaction in the original
NJL in order to investigate its effect in the quark stellar matter EoS and the structure of pure quark stars \cite{malheiro,itoh,fiolhais}.

\section{THE NJL MODEL WITH VECTOR COUPLING}

In this work we consider a two-flavor NJL Lagrangian of the form \cite{buballa1}:
\begin{align}
&{\cal {L}}=\bar{q} (i\gamma^\mu \partial_\mu - m)q +g_S[(\bar{q}q)^2+(\bar{q}i\gamma_5\overrightarrow{\tau}q)^2] \nonumber \\
&-g_V(\bar q \gamma^\mu q)^2,
\end{align}
where $q$ is a fermion field with $N_f=2$ flavors and $N_c$ colors. Apart from the bare mass $m$, the Lagrangian is chirally
symmetric $(SU(2)_L \times SU(2)_R)$. We have considered interaction terms in the scalar, pseudoscalar-isovector
and vector-isoscalar channels. The $g_S$ and $g_V$ are the scalar and vector couplings, respectively, and are assumed to be constants with dimension MeV$^{-2}$.

The mean field thermodynamic potential density at zero temperature and $\mu\geq 0$, that are the conditions inside
compact stars, with quark energy $E_p=\sqrt{p^2+M^2}$, is calculated in the Hartree approximation and is given by \cite{buballa1}:
\begin{align}\label{potential}
&\Omega (0,\mu;M,\mu_R)=-2N_fN_C \int \frac{d^3p}{(2\pi)^3} \{ E_p \nonumber \\
&+(\mu_R-E_p)\times \theta(\mu_R-E_p)\}+\frac{(M-m)^2}{4g_S} \nonumber \\
&-\frac{(\mu-\mu_R)^2}{4g_V}.
\end{align}

The thermodynamic potential density $\Omega$ depends on two parameters, the dynamical fermion mass $M$ and the
renormalized quark chemical potential $\mu_R$, which are related to the scalar $\langle \bar{\psi} \psi \rangle$ and
vector $\langle \psi^\dagger \psi\rangle$ densities at the chemical potential $\mu$ by \cite{njl1,njl2}:
\begin{equation}\label{gapmass}
M=m-2g_S\langle \bar{\psi} \psi \rangle,
\end{equation}
\begin{equation}\label{gappotencialquimico}
\mu_R=\mu-2g_V\langle \psi^\dagger \psi\rangle.
\end{equation}

Eq. (\ref{gapmass}) is the well-known NJL gap equation for the dynamical mass $M$. In general, this equation has more
than one solution, the physical being the one that minimizes $\Omega$. The vacuum contribution to the thermodynamical
potential density diverges and has to be regularized. This regularization is carried on by introducing a cutoff $\Lambda$.

Once we have the thermodynamic potential density, other thermodynamic quantities can be calculated in the standard way.
In particular the baryon number density, energy density and pressure are given respectively by:
\begin{equation}
 \rho= \frac{1}{N_c}\langle \psi^\dagger \psi\rangle = \frac{N_f}{3\pi^2}p_F^3,
\end{equation}
\begin{equation}
 \varepsilon = \Omega + \mu N_c \rho,
\end{equation}
and
\begin{equation}
 P=-\Omega,
\end{equation}
where the energy density and the pressure of the physical vacuum was set to zero. Then, for a given baryon chemical potential $\mu$, these expressions need to be evaluated for the values of $M$ and $\mu_R$ which minimize the thermodynamic potential.

\subsection{Stability of Quark Matter}

The Nambu-Jona-Lasinio (NJL) model is one of the most popular models for studying the spontaneous chiral symmetry breaking and
its restoration at finite temperatures or densities \cite{njl1,njl2}. The model contains quarks with a current quark mass which acquire a constituent quark mass $M$ through the quark interaction. In the mean field approximation, this interaction produces a quark mass, via a  condensation process of quark-antiquark pairs that spontaneously breaks the chiral symmetry. A study of the quark
matter stability was done in \cite{buballa1}, and we analyze here the pressure and energy per particle as a function of the baryonic density \cite{thesisjaziel}.
\begin{figure}[ht!]
\vspace{9pt}
\parbox[c]{6cm}{\includegraphics[width=0.5\textwidth]{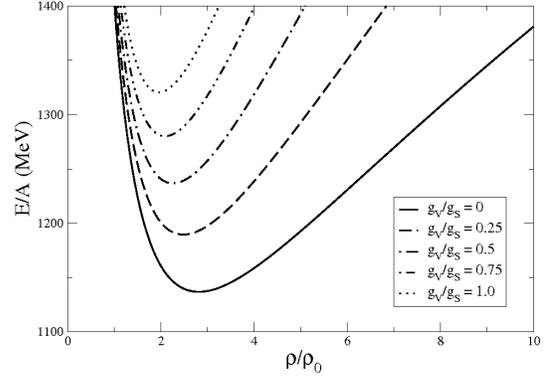}}
\vspace{-1.0cm} \caption{Energy per baryon as a function of the baryonic density $\rho/\rho_0$ for different values of the vector coupling.}
\label{enerparticle}
\vspace{-0.154cm}
\end{figure}

The character of the chiral phase transition crucially depends upon the model parameters, particularly the vector coupling constant.
For $ g_V/ g_S < 0.75$ the phase transition is of first order. However, when $g_V/g_S \geq 0.75$ it becomes of second order since for
different values of $\mu$, only the $M \neq 0$ minimum exists and it moves continuously to the $M=0$ minimum.
In the Figure \ref{enerparticle} we present the energy per particle for different values of the vector coupling. We observe that as the ratio $g_V/g_S$ increases the quark matter gets less stable.

Quark matter inside compact star is neutral with respect to electric charges and color. Moreover, the matter is in $\beta$ equilibrium,
i.e., all process of beta decay of quarks involving electrons and antineutrinos and their inverse process are
likely to occur at identical rates. Thus, we must impose both $\beta$ equilibrium and charge neutrality. Throughout 
this paper we only consider the latest stage in the star evolution, when entropy reaches the maximum and neutrinos have already diffused out.
The neutrino chemical potential is then set to zero. For $\beta$ equilibrium matter we must add the contribution of leptons to the stellar energy density and pressure. The relation between the chemical potentials of light quarks is determined by beta decay and can be
written as \cite{glendenning}:
\begin{equation}\label{eb}
 \mu_d = \mu_u + \mu_e.
\end{equation}
For charge neutrality we must impose:
\begin{equation}\label{nc}
 \rho_e =\frac{1}{3}(2\rho_u - \rho_d).
\end{equation}

Eqs. (\ref{eb}) and (\ref{nc}) need to be satisfied together with the self consistent relations for the quark effective mass during
the numerical calculation. The EoS for the leptonic sector is also needed, and is considered as a relativist degenerate Fermi gas.

\begin{figure}[ht!]
\vspace{9pt}
\parbox[c]{6cm}{\includegraphics[width=0.5\textwidth]{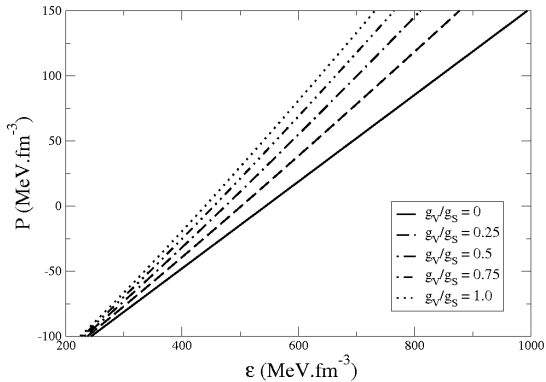}}
\vspace{-1.0cm} \caption{Pressure as a function of the energy
density for different ratios of the vector and scalar coupling
constants $g_V/g_S$.} \label{eos}
\end{figure}

The effect of the vector coupling in the quark matter EoS is shown in Figure \ref{eos}. We can see that for a certain energy
density an increase of $g_V/g_S$ increases the pressure. The equation of state EoS becomes stiffer as
expected since we know that the vector interaction is repulsive.

\section{QUARK STARS PHENOMENOLOGY}

\begin{figure}[hc!]
\vspace{9pt}
\parbox[c]{6cm}{\includegraphics[width=0.5\textwidth]{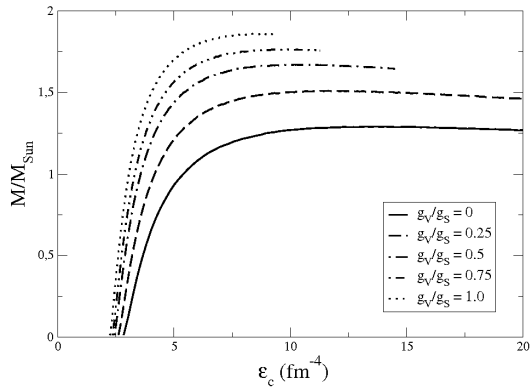}}
\vspace{-1.0cm} \caption{The star mass is plotted as a function of the
central energy density for different choices of the vector
coupling.} \label{massadensc}
\end{figure}

\begin{figure}[ht!]
\vspace{9pt}
\parbox[c]{6cm}{\includegraphics[width=0.5\textwidth]{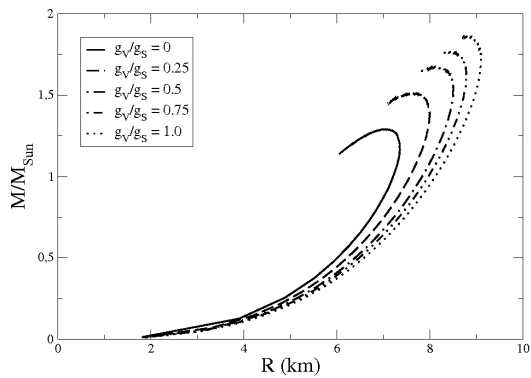}}
\vspace{-1.0cm} \caption{Mass-radius diagram obtained for four
values of the vector coupling.} \label{mr}
\end{figure}

The quark matter EoS obtained in the previous section are use to solve the Tolman-Oppenheimer-Volkoff equations. These equations are integrated from the origin, where $M(R =0)=0$ and $\varepsilon(R = 0)=\varepsilon_c$, with $\varepsilon_c$ the
central energy density, up to a point where $r = R$ and the pressure is zero. For each EoS  we obtain a unique relation between the star
mass and its central energy density \cite{glendenning,taurines,weber,bookweber}. Figure \ref{massadensc} presents the curve $M/M_\odot \times \varepsilon_c$ for
different ratios  $g_V/g_S$ as a function of the central energy density, normalized by the mass of the Sun $M_\odot = 1.9891 \times 10^{30}$kg.
As the vector coupling increases, we obtain more massive stars for the same central energy density. The reason lies in the repulsive vector coupling, which makes stiffer the EoS. The internal pressure increases for a given central energy and supports a larger star mass.
In the Figure \ref{mr} we show the $M/M_\odot \times R$ diagram for quark stars. We see that with the increase of $g_V/g_S$ the stars have
more mass and larger radii.

The masses and radii of the maximum mass stable configuration obtained with the inclusion of the vector couplings in the SU(2) version
of the NJL model are larger than the corresponding NJL without the repulsive term. The radius is still quite small, below 10 km, smaller than
the average neutron star value, 12-15 km but the mass is larger than the value 1.2 M$_\odot$ which many neutron stars
show (see Table~\ref{table:1}).
\begin{table*}[htb]
\caption{Quark star properties for the EoS described in the text.}
\label{table:1}
\newcommand{\m}{\hphantom{$-$}}
\newcommand{\cc}[1]{\multicolumn{1}{c}{#1}}
\renewcommand{\tabcolsep}{2pc} % enlarge column spacing
\renewcommand{\arraystretch}{1.2} % enlarge line spacing
\begin{tabular}{@{}lllll}
\hline
$g_V/g_S$  & \cc{$(M/M_\odot)$} & \cc{$R$(km)} & \cc{$\varepsilon_c$(fm$^{-4}$)}  \\
\hline
\m0      & \m1.29 & \m7.36 & \m13.69   \\
\m0.25   & \m1.51 & \m7.99 & \m11.61   \\
\m0.5    & \m1.67 & \m8.50 & \m11.01   \\
\m0.75   & \m1.76 & \m8.79 & \m9.56    \\
\m1.0    & \m1.86 & \m9.11 & \m8.69    \\
\hline
\end{tabular}\\[2pt]
\end{table*}

\section{CONCLUSIONS}
Our study only concerns pure quark stars with two flavors. At star surface $(P=0)$ the energy per particle is greater
than the nucleon mass and therefore within this model quark matter will hadronize. Thus, these results for pure quark stars need
to be seen with caution and we may say that it is more likely to expect in NJL model hybrid stars with a quark matter core with two and also
three quark flavors. A work in the SU(3) version of the NJL model with vector interaction is underway. A superconducting quark phase recently proposed should also be considered \cite{maneNPA,linaresBJP}.

\end{document}